\documentclass{emulateapj}

\newcommand{\lir}{\mbox{L$_{\rm IR}$}}

\newcommand{\lhcnjone}{\mbox{L$^{\prime}_{\rm HCN(1-0)}$}}
\newcommand{\lhcnjthree}{\mbox{L$^{\prime}_{\rm HCN(3-2)}$}}

\newcommand{\lmol}{\mbox{L$_{\rm mol}$}}
\newcommand{\lcojone}{\mbox{L$^{\prime}_{\rm CO(1-0)}$}}

\shorttitle{}
\shortauthors{Bussmann et al.}

\slugcomment{Accepted to ApJ Letters}

\begin{document}
\title{The Star Formation Rate - Dense Gas Relation in Galaxies as Measured by
HCN~(3-2) Emission} 

\author{R. S. Bussmann\altaffilmark{1}, D. Narayanan\altaffilmark{1,3},
Y. L. Shirley\altaffilmark{1,2}, S. Juneau\altaffilmark{1}, J.
Wu\altaffilmark{3}, P. M. Solomon\altaffilmark{4,5}, P. A. Vanden
Bout\altaffilmark{6}, J. Moustakas\altaffilmark{7},C. K. Walker\altaffilmark{1}}

\altaffiltext{1}{Steward Observatory, University of Arizona, 933 N
Cherry Ave, Tucson, AZ, 85721, USA}
\altaffiltext{2}{Bart J. Bok Fellow}
\altaffiltext{3}{Harvard-Smithsonian Center for Astrophysics, 60 Garden
Street, Cambridge, Ma, 02138, USA}
\altaffiltext{4}{Department of Physics and Astronomy, State University
of New York, Stony Brook, Stony Brook, NY, 11974, USA}
\altaffiltext{5}{Deceased}
\altaffiltext{6}{National Radio Astronomy Observatory, 520 Edgemont
Road, Charlottesville, Va, 22903, USA}
\altaffiltext{7}{Center for Cosmology and Particle Physics, New York 
University, New York, NY 10003}

\begin{abstract}

We present observations made with the 10m Heinrich Hertz Sub-Millimeter
Telescope of HCN~(3-2) emission from a sample of 30 nearby galaxies ranging in
infrared luminosity from 10$^{10} - 10^{12.5} \: L_\sun$ and HCN~(3-2)
luminosity from 10$^6 - 10^9$~K$\;$km$\:$s$^{-1}\;$pc$^2$.  We examine the
correlation between the infrared luminosity and HCN~(3-2) luminosity and find
that the best fit linear regression has a slope (in log-log space) of $0.74 \pm
0.12$.  Including recently published data from Gracia-Carpio et al. tightens
the constraints on the best-fit slope to $0.79 \pm 0.09$.  This slope below
unity suggests that the HCN~(3-2) molecular line luminosity is not linearly
tracing the amount of dense gas.  Our results are consistent with predictions
from recent theoretical models that find slopes below unity when the line
luminosity depends upon the average gas density with a power-law index greater
than a Kennicutt-Schmidt index of 1.5.  

\end{abstract}
\keywords{galaxies: evolution and ISM and starburst --- ISM: molecules --- submillimeter }

\section{Introduction}

For decades, it has been known that the star formation rate (SFR) in galaxies
is intimately related to the gas reservoir from which stars are formed
\citep{sch59}.  Observations of the galactic-averaged surface density of HI and
CO gas indicate that the SFR increases with total gas surface density (HI +
H$_2$ measured by CO~J=1-0) according to $\Sigma_{\rm SFR} \sim \Sigma_{\rm
gas}^{N}$, where $N = 1.4\pm 0.15$ \citep[Kennicutt-Schmidt law, hereafter KS
law;][]{ken98b}.  Recent studies have focused on dense molecular gas tracers
(e.g., HCN~(1-0), CO~(3-2)) and have found a tight, linear correlation between
the SFR, traced by infrared (IR) luminosity ($\lir$), and the mass of dense
gas, traced by molecular line luminosity, \lmol\ \citep{gao04a,gao04b,nar05}.
The surface density relation and the SFR-\lmol\ relation (for dense gas
tracers) appear to predict different behaviors for the underlying star
formation law in galaxies.

The linear relationship between \lir\ and HCN~(1-0) luminosity over three
decades in \lir\ found by \citet{gao04a,gao04b} in local star forming galaxies
has been interpreted as a constant star formation efficiency (SFR/M$_{\rm H2}$)
traced by dense molecular gas.  \citet{wu05} extended the observed linear
relationship between \lir\ and HCN~(1-0) luminosity to Galactic clumps,
positing that if HCN~(1-0) emission faithfully traces dense molecular core mass
(above a cutoff luminosity of $L_{\rm bol} > 10^{4.5} \: L_\sun$), then
constant SFR per unit mass is a result of a dense molecular clump comprising a
``fundamental unit'' of star formation.  The observed extragalactic linear
correlation is a natural extension of the constant SFR per unit mass observed
toward dense molecular clumps in the Milky Way \citep{plu92,shi03,shi07}.  In
this picture, ultraluminous IR galaxies such as Arp~220 which lie on the
linear SFR-HCN~(1-0) relation simply contain more cluster forming units, and a
higher fraction of dense molecular gas. 

The interpretation outlined in \citet{wu05} predicts a linear relation between
SFR and tracers of even higher critical ($n_{\rm crit}$) density than
HCN~(1-0).  However, recent theoretical models from \citet[][hereafter
N07]{nar07} and \citet[][hereafter KT07]{kru07} predict that the power-law
index between the SFR and higher critical density tracers such as HCN~(3-2)
should in fact be below unity.  The physical explanation for this behavior is
that in systems with predominantly low density gas, emission from high critical
density lines originates in the extreme tails of the density distribution
resulting in a L$_{\rm mol}$-$<$$n$$>$ relation with a slope greater than
unity.  This effect drives a SFR-L$_{\rm mol}$ relation with a slope below
unity for tracers with higher $n_{\rm crit}$ than that of HCN (J=1-0).  To
test this prediction, we have measured the HCN~(3-2) line luminosity
($\lhcnjthree$) from a sample of 30 galaxies and compared our results with
recently published data from \citet[][hereafter GC07]{graciacarpio2007}.  Both
datasets show a \lir\ - \lhcnjthree\ slope that is significantly below unity,
in agreement with the model predictions.  


\section{Observations}\label{sec:obs}

 Observations of HCN~(3-2) ($\nu_{rest} = 265.886431 \: {\rm GHz}$) were
obtained from 2007 February through 2007 June using the 10m Heinrich Hertz
Sub-millimeter Telescope (SMT) on Mt. Graham (Arizona).  Our sample includes 30
galaxies and covers a broad range of IR luminosities: $L_{\rm IR} \sim
10^{10.2} - 10^{12.5} \: L_\sun$.  Central positions of all galaxies were
observed in HCN~(3-2).  One nearby galaxy ($D < 7 \:$Mpc) was mapped (NGC0253).
The FWHM of the SMT at 265~GHz is $\sim$30$\arcsec$, such that a single beam
covers the central kpc of galaxies beyond 7~Mpc.  

The observations were made using the new ALMA sideband separating receiver and
the Forbes Filterbank spectrometer.  This dual-polarization receiver uses
image-separating Superconductor-Insulator-Superconductor (SIS) mixers that are
significantly more sensitive than conventional receiver systems using
quasioptical techniques for image separation.  Receiver temperatures were
typically $\sim 100 \:$K and system temperatures ranged from 200 - 400~K.  The
2048 channel, 1~MHz resolution spectometer was split into two 1~GHz bandwidths,
one for each polarization, corresponding to a velocity coverage
per-polarization of 1130~km$\:$s$^{-1}$.  All of our targets were observed in
beam-switching mode with a chop rate of 2.2~Hz and a beam throw of 2$\arcmin$.
We employed a 6~GHz intermediate frequency with the HCN~(3-2) line in the lower
side band.  Image rejections of the upper side band were typically 18-20~dB.

Our observing strategy involved pointing and calibration observations
approximately every two hours using either Jupiter or Saturn when available,
otherwise DR21 or W3OH.  Calibration scans were obtained in position-switch
mode with reference off positions of 5$\arcmin$ (30$\arcmin$ for DR21) in right
ascension.  We found typical pointing errors of 2-3$\arcsec$ and measured the
main beam efficiency, $\eta_{\rm mb}$, to be $0.67 \pm 0.04$ for filterbank A
(H-polarization) and $0.80 \pm 0.04$ for filterbank B (V-polarization).
Antenna temperatures are converted to main-beam temperatures using $T_{\rm mb}
= T_A^* / \eta_{\rm mb}$.  We assumed a systematic calibration uncertainty of
$\sim$20\%.  The integrated intensity of HCN~(3-2) emission was calculated
using velocity intervals based on the HCN~(1-0) line profiles.  We computed the
HCN line luminosity using Equation~1 in \citet{gao04b} for objects further than
7~Mpc away and using their Equation~4 for NGC0253, the nearby galaxy which we
mapped.  Our mapping strategy involved 5 pointings along the major axis and 3
along the minor axis, each separated by 15$\arcsec$.  Our results are presented
in Table~\ref{tab:results}.

\begin{deluxetable}{lccc}
\tabletypesize{\small}
\tablecolumns{4}
\tablewidth{230pt}
\tablecaption{SMT HCN(3-2) Data}
\tablehead{
\colhead{} & \colhead{$I_{\rm HCN(3-2)}$} & \colhead{$L^\prime_{\rm HCN(3-2)}$} &
\colhead{$\lir$} \\
\colhead{Source} & \colhead{(K~km$\:$s$^{-1}$)} & 
\colhead{(10$^6$~K~km$\:$s$^{-1}$~pc$^2$)} & \colhead{(10$^{11}$~$L_\sun$)}
}

\startdata
ARP193 			& 0.36$\pm$0.11 & 82$\pm$25 & 5.1$\pm$1.0 \\
ARP220 			&4.51$\pm$0.90 & 634$\pm$127 & 18.3$\pm$3.7  \\
ARP55 			& $<$ 0.23 &  $<$ 147 & 5.2$\pm$1.0  \\
IC342\tablenotemark{a}	& 1.76$\pm$0.37 & 0.25$\pm$0.05 & 0.13$\pm$0.03 \\
IRAS10565 		&   $<$ 0.19 &  $<$ 148 & 11.0$\pm$2.2                       \\
IRAS17208 		&   $<$ 0.27 &  $<$ 203 & 29.1$\pm$5.8                       \\
IRAS23365 		&   $<$ 0.18 &  $<$ 293 & 14.2$\pm$2.8                       \\
M82\tablenotemark{a}  	& 6.21$\pm$1.26 & 2.7$\pm$0.6 & 0.53$\pm$0.11     \\
MRK231  		& $<$ 0.15 &  $<$ 108 & 26.5$\pm$5.3                               \\
MRK273  		& $<$ 0.13 &  $<$ 77 & 15.6$\pm$3.1                               \\
NGC0253\tablenotemark{b}  & 13.67$\pm$2.85 & 8$\pm$2 & 0.25$\pm$0.05  \\
NGC0520  		& 0.55$\pm$0.13 & 14$\pm$3 & 0.8$\pm$0.2                   \\
NGC0660  		& 1.19$\pm$0.29 & 4$\pm$1 & 0.31$\pm$0.06                   \\
NGC0695  		&   $<$ 0.13 &  $<$ 56 & 4.0$\pm$0.8                              \\
NGC1068 	 	& 6.01$\pm$1.22 & 38$\pm$8 & 1.0$\pm$0.2                   \\
NGC1614  		&   $<$ 0.21 &  $<$ 23 & 3.5$\pm$0.7                              \\
NGC2146  		& 0.68$\pm$0.15 & 2.6$\pm$0.6 & 1.1$\pm$0.2                 \\
NGC2903  		& 0.61$\pm$0.16 & 0.9$\pm$0.2 & 0.15$\pm$0.03                 \\
NGC3079  		& 2.58$\pm$0.53 & 16$\pm$3 & 0.6$\pm$0.1                   \\
NGC3628  		& 1.04$\pm$0.22 & 3.6$\pm$0.8 & 0.18$\pm$0.04                 \\
NGC3690  		& 0.83$\pm$0.19 & 39$\pm$9 & 7.1$\pm$1.4                   \\
NGC3893  		&   $<$ 0.23 &  $<$ 1 & 0.16$\pm$0.03                              \\
NGC4414  		&   $<$ 0.19 &  $<$ 0.5 & 0.35$\pm$0.07                             \\
NGC6240  & 0.88$\pm$0.19 & 225$\pm$49 & 7.0$\pm$1.4                   \\
NGC6701  & 0.36$\pm$0.10 & 27$\pm$8 & 1.1$\pm$0.2                   \\
NGC7331  &   $<$ 0.16 &  $<$ 0.5 & 0.37$\pm$0.07                             \\
NGC7469  & 0.57$\pm$0.14 & 65$\pm$16 & 3.3$\pm$0.6                   \\
NGC7771  & 0.75$\pm$0.17 & 66$\pm$15 & 2.2$\pm$0.4                   \\
UGC051017  &   $<$ 0.11 &  $<$ 74 & 9.6$\pm$1.9                             \\
VIIZW31  &   $<$ 0.32 &  $<$ 373 & 9.0$\pm$1.8                              \\
\enddata
\tablenotetext{a}{~ lower limits due to spatial undersampling}
\tablenotetext{b}{~ mapped}
\label{tab:results}
\end{deluxetable}

\section{Results}\label{sec:results}

\begin{figure}[!tbp]
\epsscale{1.00}
\plotone{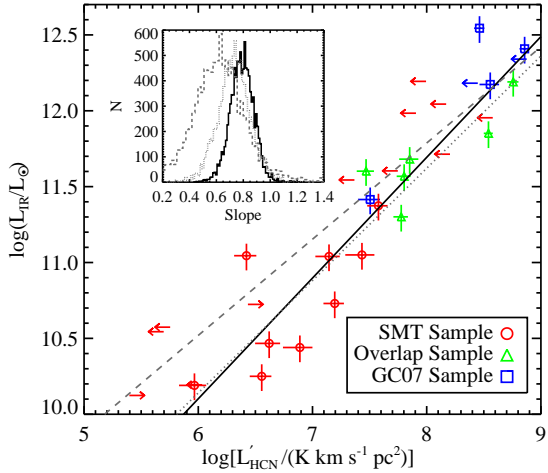}

\caption{ \lir\ as a function of \lhcnjthree\ for the sample of
extra-galactic sources reported here (open circles), those observed by
GC07 (blue squares), and those detected by both surveys
(green triangles).  Objects not detected above the 2$\sigma$ level are shown as
upper limits while sources within 7~Mpc that were not mapped (IC342 and M82)
are shown as lower limits.  Error bars represent 1$\sigma$ uncertainties.  The
solid line is a best fit linear regression to the galaxies with detections from
the combined sample and has a slope of 0.79.  The dotted and dashed lines
represent the same quantities but as applied to the SMT sample only and
GC07 sample only, respectively.  The inset shows the
distribution of slope values returned by the fitting routine for each of these
three samples.  In the combined sample, a slope of unity
is ruled out at the 99\% confidence level.    \label{figure:irhcn}}

\end{figure}

In Figure~\ref{figure:irhcn}, we show the \lir\ - \lhcnjthree\ relation using
the data from Table~\ref{tab:results} and from GC07.  
Although in principle both star formation and Active Galactic Nuclei (AGN)
processes heat the dust in a galaxy, the IR luminosity from our sample of
nearby galaxies is largely uncontaminated by AGN, and so we use \lir\ as a
proxy for the SFR.  Some exceptions are two objects presented in this paper
(NGC1068 and NGC7469) as well as two objects from \citet[Mrk231 and
Mrk273;][]{graciacarpio2007}.  Excluding these sources from the fit does not
significantly alter our results (a more detailed investigation into the role of
AGN will be presented in a companion paper; S.  Juneau et al.  in prep).
Furthermore, these results (particularly our best-fit slope values) do not
change significantly if one extrapolates from the FIR luminosity to estimate
$\lir$.
We conservatively assume a 20\% uncertainty in the correction factor needed to
generate a \lir\ value from the {\it IRAS} flux densities \citep[Tab.~1
in][]{sanders96}, which ends up dominating the total error budget.  Adjusting
this uncertainty from 10\% to 30\% does not have a significant effect on the
resulting best-fit slope.  

We use the publically available Bayesian Monte Carlo Markov Chain routines of
\citet{kel07} to compute the linear regression between log($\lir$) and
log($\lhcnjthree$).  This routine  assumes that the distribution of the
independent variable can be well described by a mixture of Gaussian functions
and accounts for heteroscedastic errors in both \lir\ and $\lhcnjthree$. The
posterior distributions of possible slopes and y-intercepts are sampled.  We
define the best fit using the median slope and intercept values.  The dotted
grey line in Figure~\ref{figure:irhcn} is the best fit to those galaxies
observed by the SMT with $>$2$\sigma$ detections (excluding IC342 and M82, for
which we have only lower limits on $\lhcnjthree$) and is described by the
following equation:

\begin{equation}
\log L_{\rm IR} = (0.74 \pm 0.12) \times \log L_{\rm HCN~(3-2)} + (5.7 \pm
0.9).
\end{equation} 

\noindent A slope of unity is inconsistent with the data at the 98\% confidence
level.  However, we caution that the formal uncertainty is large enough to
encompass a slope of unity at the 2$\sigma$ level.  We note that using a simple
least squares fit routine (SVDFIT in IDL) instead of the linear regression code
from \citet{kel07} produces a slope of 0.73.  

We can expand our sample by including data from GC07.  These authors observed
HCN~(3-2) emission from a sample of 13 galaxies with the IRAM 30m telescope, of
which 10 were significantly detected.  We observed 9 of these sources and were
able to detect 6 of them.  
Combining our sample with that from GC07, we can place an even stronger
constraint on the value of the slope.  For the sources marked in green
triangles, we use the average \lhcnjthree\ value of the two surveys
(measurements of the overlapping objects were in agreement to within 20\%), but
the \lir\ value as given in GC07, since they use a more accurate prescription
for computing the total IR luminosity.  

We note that a similar sample of nearby galaxies has been observed by
\citet{pag97}.  Unfortunately, a direct comparison between their sample and
ours is problematic because they employed a dual sideband receiver system and
did not measure the sideband gain ratio, which can vary by a factor of two.  We
examined the integrated intensities of the seven galaxies appearing in both
samples and found the values to be consistent within this factor of two.
HCN~(3-2) luminosities have been published for the nuclear region of a set of
nearby galaxies by \citet{krips07}.  However, these observations were conducted
with the IRAM 30m telescope (with a beam-size a factor of 3 smaller than the
SMT) and therefore provide only lower limits to the full \lhcnjthree\ emission.
For the purposes of this paper, the best comparison sample is that of GC07, so
we place our focus there for the remainder of the discussion.

We compute the distribution of best-fitting slopes when considering only the
published detections in GC07.  This is shown with the grey
dashed line in the inset of Figure~\ref{figure:irhcn}, where the median and
standard deviation are $0.63 \pm 0.20$.  The best fit linear regression is
shown with the dashed line.  The next step we take is to combine all available
data and recompute the best fit slope and y-intercept.  This results in a
narrower distribution of slope values with a median below unity: $0.79 \pm
0.09$.  The distribution is shown as the solid black line in the inset of
Figure~\ref{figure:irhcn}, and the best fit line is shown in the full plot.
Using the full, combined dataset, a slope of unity is ruled out at the 99\%
confidence level (the median value is 2.5$\sigma$ from unity).   Using IDL's
SVDFIT, we find a slope of 0.79 using the full combined dataset.  One of the
objects in our sample (NGC2146) lies significantly off the best fit relation,
in the sense of either a lower \lhcnjthree\ value or a greater $\lir$.  To
explore the extent to which this affects the resultant best fit slope,  we
remove this object from the sample and re-compute the slope, finding a larger
slope with a smaller dispersion: 0.84$\pm$0.07.  This remains signicantly below
unity and within one standard deviation of the result using the full sample.
Finally, GC07 find evidence for a change in the \lir\ -
\lhcnjone\ relation at $L_{\rm IR} \sim 10^{11} \: L_\sun$.  Restricting our
sample to galaxies above this IR luminosity, we find a slightly shallower best
fit slope of 0.64$\pm$0.13.  

\section{Discussion}\label{sec:disc}

The sub-unity slope observed in the \lir\ - \lhcnjthree\ relation presents a
challenge to our understanding of the molecular star formation rate law.  If
the constant SFR per unit dense gas mass applies to observations of HCN~(3-2)
emission, then a linear relationship between \lir\ and \lhcnjthree\ should be
observed.  One underlying assumption is that the HCN~(3-2) molecular line
luminosity is linearly tracing the dense gas mass.  This may not be valid when
observing unresolved galaxies that include large quantities of sub-thermally
excited gas.  Recent theoretical models from N07 and KT07 suggest that the
index between the SFR and higher critical density tracers such as HCN~(3-2)
should in fact be below unity.  N07 used hydrodynamical simulations of isolated
galaxies and equal-mass galaxy mergers coupled with a 3D non-local
thermodynamic equilibrium radiative transfer code to probe the relationship of
the dense molecular gas phase in galaxies with HCN and CO emission across a
variety of rotational transitions.  Meanwhile, KT07 use escape probability
radiative transfer simulations coupled with models of turbulence-regulated
giant molecular clouds.  While their work did not specifically explore the
potential relations between SFR and higher critical density lines such as
HCN~(3-2), fundamentally their conclusion behind the physical driver of the
$\lir$-\lhcnjone\ relation is similar to that of N07, and thus similar results
for higher lying transitions are expected.

According to N07 and KT07, the SFR-HCN~($u$-$l$) index is parameterized in
terms of how the molecular line luminosity, \lmol\ is related to the mean
molecular gas density, $<$$n$$>$, of a given galaxy. If \lmol\ grows as
$<$$n$$>$, the SFR-\lmol\ relation will have an index of $\sim$1.5 (e.g. the
observed $\lir$-$\lcojone$ index).  On the other hand, a \lmol-$<$$n$$>$
power-law index equal to the KS index produces a linear SFR-\lmol \
relationship.  In this picture, the fundamental relationship is the volumetric
version of the KS relation, and the observed linear SFR-HCN~(1-0) relationship
in galaxies results from the HCN~(1-0) luminosity on average being related to
the mean molecular gas density by an index similar to the KS index (i.e.
L$\sim$$<$$n$$>$$^{1.5}$).  Since \lhcnjone\ rises linearly with $\lir$,
HCN~(1-0) emission is a useful proxy for the total SFR, once the effects of AGN
have been properly taken into account.

\begin{figure}[!tbp]
\includegraphics[scale=0.35,angle=90]{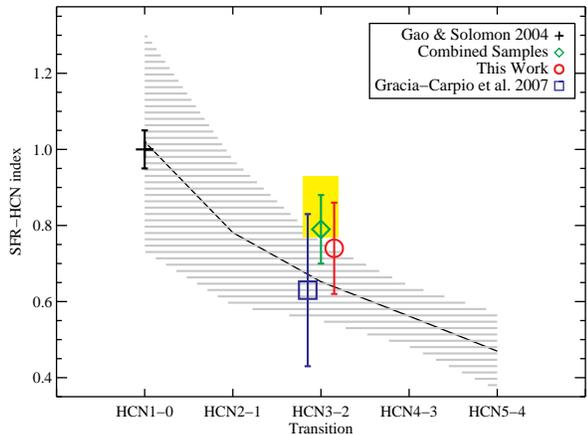}

\caption{ Predicted slopes in log($\lir$) - log($\lhcnjthree$) space as a
function of $J$ transition of HCN, as predicted from theoretical models by N07
(grey horizontal lines) and KT07 (yellow shaded area at the HCN~(3-2)
transition).  Observational constraints on the slope measured from HCN~(1-0)
and HCN~(3-2) emission are shown with their error bars.  The best-fit slopes
are consistent with the model predictions from both N07 and KT07.
\label{figure:slopes}}

\end{figure}

Alternatively, if the \lmol-$<$$n$$>$ power-law index is greater than the KS
index, the SFR-\lmol\ index will be below unity.  Physically, this can be
understood as resulting from high critical density lines originating in extreme
tails of the density distribution and causing a steepening of the
$\lmol$-$<$$n$$>$ power law index compared to lower critical density lines.
N07 predict that the SFR will be related to the HCN~(3-2) luminosity from
galaxies by an index of $\sim$0.7.  Figure~\ref{figure:slopes} is a
reproduction of Figure~8 in N07, but with new constraints our data---as well as
the data from GC07---place on the SFR-HCN power law index.  The scatter in the
predicted slope is represented by the horizontal grey lines and is computed by
randomly drawing a sample of 19 galaxies (which is the size of our combined
dataset) out of a set of 100 model galaxies 1000 times.  The yellow shaded
region shows the range of expected results from KT07 utilizing their publically
available code with parameters appropriate for ``normal'', ``intermediate'',
and ``starburst'' galaxies.  In the KT model, the SFR-HCN~(3-2) relation is
predicted to have a slope below unity in the range 0.77 - 0.93.  The plotting
symbols show the best-fit and 1$\sigma$ range in slopes for the datasets
presented here as well as that of GC07 and \citet{gao04b}.  All HCN~(3-2)
measurements are consistent with the models shown here.  

It is important to note that we do not account for potential chemistry-related
issues.  While some evidence may exist for HCN-related chemistry in the
vicinity of the hard X-ray flux associated with an AGN, both theoretical and
observational investigations have found mixed results \citep{lin06, gra06,
mei07,com07}.  Studies of the nuclear region of galaxies like that done by
\citet{krips07b} of the Seyfert~2 galaxy NGC6951 will be essential to
understanding all of the potentially relevant chemical processes.  Finally, our
sample includes only nearby galaxies; the most distant objects in our sample
lie roughly 100~Mpc away (Arp193 and NGC6240).  Among others, \citet{gao07} and
\citet{rie07} have found evidence that the \lir\ - \lhcnjthree\ relation
steepens at high redshift and/or high $\lir$.

\section{Conclusions}\label{sec:conc} We present observations of HCN~(3-2)
emission from a sample of 30 nearby galaxies ranging in IR luminosity from
10$^{10} - 10^{12.5} \: L_\sun$ and HCN~(3-2) luminosity from 10$^6 -
10^9$~K$\:$km$\:$s$^{-1}\:$pc$^2$.  We find a best-fit slope in
log($\lir$)-log($\lhcnjthree$) space of 0.74 and exclude a slope of unity at
the 98\% confidence level (although the formal uncertainty is large enough to
include a slope of unity at the 2$\sigma$ level).  Adding data recently
published in the literature yields a slope of 0.79 and tightens the
distribution of possible slopes such that a slope of unity is excluded at the
99\% confidence level for this sample of galaxies.  This sub-unity slope may be
an indication that the HCN~(3-2) molecular line luminosity is not linearly
tracing the dense gas.  Our results are consistent with predictions from recent
theoretical models by N07 and KT07, who predict slopes less than unity when the
line luminosity - average gas density relation has a power-law index greater
than the KS index.  We wish to emphasize that the results shown here are
pertinent to HCN~(3-2) molecular line emission and do not contradict results
from previous efforts showing a tight, linear correlation between \lir\ and
$\lhcnjone$.  Indeed, the models discussed in this paper successfully account
for this behavior as well.


\acknowledgements This work benefited from helpful conversations with Mark
Krumholz, Joop Schaye and Todd Thompson. We thank the anonymous referee for
their useful comments that helped improve the paper, as well as the Arizona
Radio Observatory for their support throughout these observations.

\end{document}